\documentstyle[epsfig,mncite]{mn}

\begin{document}
\LARGE \normalsize \title{KiloHertz quasi--periodic oscillations
difference frequency exceeds inferred spin frequency in 4U~1636--53}

\author[P. G. Jonker et al.]
{Peter G. Jonker$^1$ \thanks{email : peterj@ast.cam.ac.uk}, 
M. M\'endez$^2$,
M. van der Klis$^3$\\
$^1$Institute of Astronomy, Cambridge University,
Madingley Road, CB3 0HA, Cambridge; \\
$^2$SRON, National Institute for Space Research;
m.mendez@sron.nl\\ 
$^3$Astronomical Institute ``Anton Pannekoek'',
University of Amsterdam, and Center for High-Energy Astrophysics,
Kruislaan 403, 1098 SJ \\ 
Amsterdam; michiel@astro.uva.nl\\
}
\maketitle

\begin{abstract}
\noindent
Recent observations of the low--mass X--ray binary 4U~1636--53 with
the {\it Rossi X--ray Timing Explorer} show, for the first time, a
kiloHertz quasi--periodic oscillation (kHz QPO) peak separation that
exceeds the neutron star spin frequency as inferred from burst
oscillations. This strongly challenges the sonic--point beat frequency
model for the kHz QPOs found in low--mass X--ray binaries. We detect
two simultaneous kHz QPOs with a frequency separation of
323.3$\pm$4.3~Hz in an average Fourier power spectrum of observations
obtained in September 2001 and January 2002. The lower kHz QPO
frequency varied between 644~Hz and 769~Hz. In previous observations
of this source the peak separation frequency was $\sim$250~Hz when the
lower kHz QPO frequency was $\sim$900~Hz.  Burst oscillations occur in
4U~1636--53 at $\sim$581~Hz and possibly at half that frequency
(290.5~Hz). This is the first source where the peak separation
frequency is observed to change from less than (half) the burst
oscillation frequency to more than that. This observation contradicts
all previously formulated implementations of the sonic--point beat
frequency model except those where the disk in 4U~1636--53 switches
from prograde to retrograde. 

\end{abstract}


\section{Introduction}
\label{intro}
\noindent
In recent years kiloHertz quasi--periodic oscillations (kHz QPOs) have
been found in power spectra of more than 20 low--mass X--ray binaries
(LMXBs) using observations made with the {\it Rossi X--ray Timing
Explorer (RXTE)} satellite (\pcite{va2000}). In most sources the kHz
QPOs are found in pairs. The high frequencies of these QPOs suggest
that at least one of the two peaks reflects the orbital motion of
matter close to the neutron star. Another high frequency phenomenon
was detected with {\it RXTE}: during some type I X--ray bursts nearly
coherent oscillations which slightly drift in frequency, the
so--called burst oscillations, have been observed in the power spectra
of ten sources (for a review see \pcite{2001AdSpR..28..511S}). The
X--ray flux--modulation is consistent with being due to the changing
aspect of a drifting hot spot on the surface of the neutron
star. Therefore, these burst oscillations are thought to reflect the
spin frequency of the neutron star. In high luminosity `Z'--source
LMXBs QPOs at low frequencies were found with the {\it EXOSAT} and
{\it GINGA} satellites (for a review see
\pcite{1989ARA&A..27..517V}). Observations with the {\it RXTE}
satellite have also revealed low--frequency QPOs in a number of atoll
sources (e.g. \pcite{1998ApJ...499L..41H}; \pcite{fova1998})

The atoll source 4U~1636--53 is one of the few sources that displays
all of the high--frequency QPO phenomena that have been observed in
LMXBs; two kHz QPOs have been observed (\pcite{1996ApJ...469L..17Z};
\pcite{wivava1997}) as well as a sideband to the lower of the two kHz
QPOs (\pcite{2000ApJ...540L..29J}) and burst oscillations at $\sim$581
Hz (\pcite{1997IAUC.6541....1Z}; \pcite{stzhsw1998}). \scite{mi1999}
presented evidence that these oscillations are in fact the second
harmonic of the neutron star spin of $\sim$290.5 Hz. However, using
another dataset these findings were not confirmed
(\pcite{2001AdSpR..28..511S}).

Since the frequency difference between the two kHz QPO peaks (the peak
separation, $\Delta\nu$) is nearly equal to (half) the burst
oscillation frequency a beat frequency mechanism was proposed for the
kHz QPOs (\pcite{stzhsw1996}). In such a model the higher frequency
QPO is attributed to the orbital frequency of clumps of plasma at a
special radius near the neutron star while the lower frequency QPO is
due to the beat between the orbital frequency and the stellar spin
frequency. A specific model incorporating the beat frequency
mechanism, the sonic--point beat frequency model, is now one of the
leading models explaining kHz QPOs in LMXBs
(\pcite{milaps1998}). However, subsequent findings complicated the
picture. As was first shown in Sco~X--1, the peak separation decreases
as the kHz QPO frequencies increase (\pcite{vawiho1997}). Furthermore,
in 4U~1636--53 the peak separation was found to be less than half the
burst oscillation frequency (\pcite{mevava1998b}). These findings
could be explained within the sonic point beat frequency model
(\pcite{2001ApJ...554.1210L}) by taking into account the effect of the
inward velocity of the plasma on the emerging kHz QPO frequencies. A
firm prediction of this revised sonic--point beat frequency model is
that for prograde spinning accretion disks the kHz QPO peak separation
is smaller than the neutron star spin frequency. This follows from the
detailed description of the plasma flow patterns in this model which
produce a number of corrections to the observed frequencies relative
to the true orbital and beat frequency, all of which are expected to
make $\Delta\nu$ smaller (\pcite{2001ApJ...554.1210L}). Indeed, until
now, in various sources the peak separation was always found to be
consistent with, or less than (half) the burst oscillation frequency.

In this {\it Letter} we show that in recent observations of
4U~1636--53 the kHz QPO peak separation is significantly larger than
half the burst oscillation frequency. Using other observations
\scite{mevava1998b} had shown earlier that the peak separation was
significantly smaller than half the burst oscillation frequency. This
means that in order for the sonic--point beat frequency to explain
both observations further changes will have to be made in the
description of the flow pattern. The only change in the flow pattern
previously described that could produce the observed change in the
separation frequency is that in which the accretion disk changes from
prograde to retrograde when the lower kHz QPO moves from $\sim$750 Hz
to $\sim$800 Hz.

\section{Observations and analysis}
\label{analysis}
\noindent
4U~1636--53 was observed in September 2001 and January 2002 with the
proportional counter array (PCA; \pcite{jaswgi1996}) onboard the {\it
RXTE} satellite (\pcite{brrosw1993}). Due to the reduced duty cycle of
the PCA detectors only a subset of the 5 detectors is operational for
most of the time. In this analysis we only used data in which 4 of the
5 detectors were operational. A log of the observations is given in
Table~\ref{log}. In total $\sim$28.4 ksec of data were used. Data were
obtained using an Event--mode in which information on detected X--ray
photons is stored onboard and telemetered to the ground on an
event--to--event basis. The time resolution of the data is
$\sim$125$\mu$s and {\it RXTE}'s effective energy range (2--60 keV) is
covered by 64 channels.

\begin{table*}
\caption{Log of the observations used in this analysis.}
\label{log}
\begin{center}
\begin{tabular}{cccc}
\hline

Observation ID  & Used segments & Amount of good data (ksec) & Count
rate$^a$ cnt/s/PCU\\
\hline
\hline
60032-01-09-00 & 1     & 3.3 & 230  \\
60032-01-09-03 & all   & 2.5 & 220  \\
60032-01-19-00 & 1, 2, 3  & 8.6 & 205  \\
60032-01-21-00 & 7, 8     & 6.7 & 180  \\
60032-01-22-00 & 1, 2 & 7.3 & 180 \\

\end{tabular}
\end{center}
{\footnotesize$^a$ Average background subtracted count rate per PCU
(2--60 keV)}
\end{table*}

Using this data we calculated power density spectra of segments of
64~s up to a Nyquist frequency of 4096 Hz in one energy band covering
the total PCA energy range. We used a dynamical power spectrum
displaying consecutive power spectra rebinned to a frequency
resolution of 2 Hz and a time resolution of 128~s to visualize the
time evolution of the lower kHz QPO for all observations except for
the power spectra of observation 60032-01-21-00 which were rebinned to
a time resolution of 162s (segment 7) and 256s (segment 8), and the
power spectra of observation 60032-01-22-00, which were treated
separately. We traced the frequency of the lower kHz QPO and fitted
the power spectrum in a window of 200 Hz centered on the traced
frequency with a Lorentzian and a constant. All peaks in each power
spectrum of 128~s, 162~s, and 256~s had a significance of
$\sim3\sigma$ or more. The frequency of the bin associated with the
peak of the fitted Lorentzian was taken to be the frequency of the QPO
in that power density spectrum. Finally, we used the shift--and--add
method described by \scite{mevava1998b} to shift the lower kHz QPO to
the same (arbitrary) frequency in each power spectrum before
averaging. This assures that at least for timescales longer than 256~s
the determined peak separation is not influenced by the fact that the
two kHz QPOs have a different frequency--amplitude relation
(\pcite{2001ApJ...561.1016M}).

We averaged the aligned power spectra and fitted the average power
spectrum from 256--1500 Hz with a constant, to represent the power
introduced by the Poisson counting noise, and two Lorentzians to
represent the two kHz QPOs. Errors on the fit parameters were
determined using $\Delta\chi^2 = 1.0$ ($1\sigma$ single
parameter). The observation 60032-01-22-00 was treated separately
since we were not able to find a more than 3$\sigma$ lower kHz QPO
using data stretches shorter than or equal to 256~s. Instead we
averaged the power spectra in this observation without shifting the
data. We note that by doing so we may have artificially introduced
deviations in the kHz QPO peak separation if the two kHz QPOs moved in
frequency during the observation for reasons outlined above.

\begin{figure*}
\leavevmode{\psfig{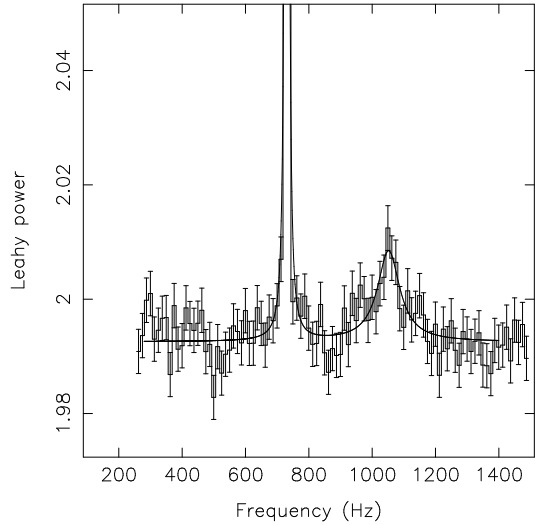}}\caption{ Average 
power density spectrum combining all shifted data used in
this paper. The peak of the lower kHz QPO is off the vertical
scale. The kHz QPO frequency separation is 323.3$\pm$4.3~Hz. The power
spectra have been shifted before averaging, therefore, only the
frequency difference between the peaks is meaningful.}
\label{total}
\end{figure*}

\begin{figure*}
\leavevmode{\psfig{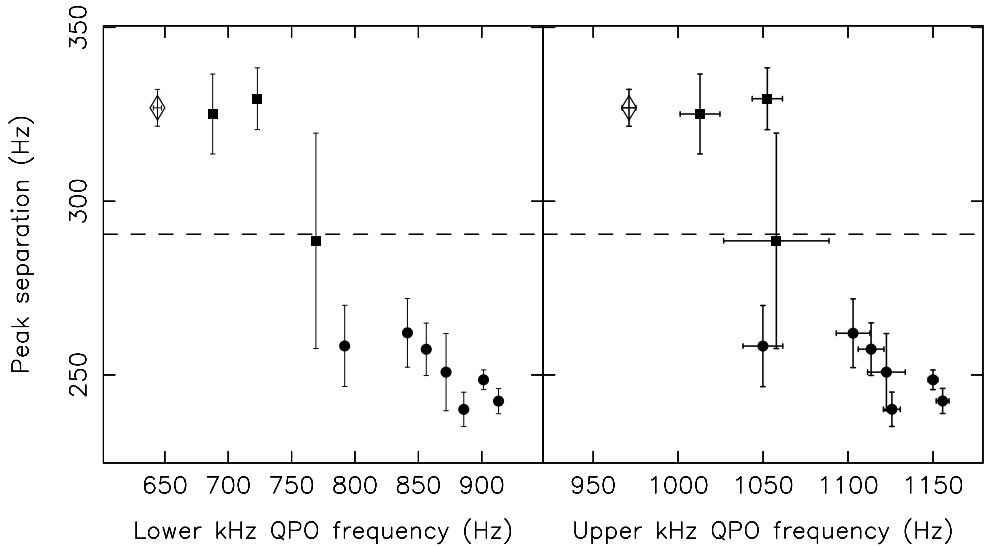}}\caption{
{\it Left:} The kHz QPO frequency separation as a function of the
frequency of the lower kHz QPO. {\it Right:} The kHz QPO frequency
separation as a function of the frequency of the upper kHz QPO. The
frequency of half the burst oscillation frequency (290.5 Hz) is
indicated by the dashed line. The peak separation varies from
significantly more than half the burst oscillation frequency [diamond
(unshifted data) and squares (shifted data); this work] to
significantly less than that (dots; Di Salvo et al. 2002; see also
Wijnands et al. 1997 and M\'endez et al. 1998). Error bars on the
frequency of the lower kHz QPO are generally smaller than the size of
the symbols.}
\label{delta}
\end{figure*}

The fits of the average power spectra were good, with a reduced
$\chi^2$ slightly larger than 1 for 146 degrees of freedom.  We found
that, combining all observations, except 60032-01-22-00, the frequency
separation between the two kHz QPO peaks was 323.3$\pm$4.3~Hz,
significantly larger than 290.5~Hz the putative spin frequency
according to an analysis of \scite{mi1999}, but still significantly
smaller than 581 Hz, the frequency of the strong burst oscillations
found by \scite{1997IAUC.6541....1Z} and \scite{stzhsw1998} (see
Figure~\ref{total}). We then divided the data set into three parts
according to the frequency of the lower kHz QPO and measured the peak
separation in each of the parts. For each part we also averaged the
power spectra {\it without} shifting in order to measure the average
frequency of the lower kHz QPO and that of a low--frequency QPO
changing in frequency from $\sim$24--42 Hz. In Table~\ref{meas} we
present our results on the kHz QPOs. The fractional rms amplitude of
the low--frequency QPO changed from 7.8$\pm$0.6 per cent to
4.3$\pm$0.7 per cent (2--60 keV) while its frequency increased from
23.6$\pm$2.4 Hz to 42$\pm$4 Hz and its FWHM was consistent with being
constant at 26$\pm$9 Hz. In Figure~\ref{delta} we plot the measured
peak separation ($\Delta\nu$) as a function of the frequency of the
lower and upper kHz QPO ({\it left} and {\it right} panel,
respectively). The squares are our new measurements using the
shift--and--add method, the diamond is our measurement using
observation 60032-01-22-00, and the dots are the measurements of
\scite{disalvoinprep} based on {\it RXTE} data obtained from
1996--2000 (see also \pcite{wivava1997};
\pcite{mevava1998b}). Clearly, $\Delta\nu$ changes from more than half
the burst oscillation frequency to less than that. The average
$\Delta\nu$ of the three leftmost points is 327.2$\pm$4.2 Hz and that
of the dots is 247.1$\pm$1.9 Hz making the jump in
$\Delta\nu\,70.1\pm4.6$ Hz.

\begin{table*}
\caption{The kHz QPO frequency separation ($\Delta\nu$), the
frequency, the fractional rms amplitude (2--60 keV), and the
full--width--at--half--maximum (FWHM) of the lower kHz QPO
($\nu_{lower~kHz}$), the fractional rms amplitude (2--60 keV) and the
FWHM of the upper kHz QPO ($\nu_{upper~kHz}$).}
\label{meas}
\begin{center}
\begin{tabular}{cccccccc}
\hline
 & Lower~kHz QPO & & & & Upper~kHz QPO & & Peak separation \\
\hline
\hline
rms (\%) & FWHM (Hz) & $\nu_{lower~kHz}$ (Hz)$^a$ & & rms (\%) &
FWHM (Hz) & & $\Delta\nu$ (Hz) \\
\hline

7.7$\pm$0.4 & 52$\pm$8 & 644.2$\pm$3.2 & & 6.9$\pm$0.5 & 52$\pm$11 & &
326.9$\pm$5.3  \\
7.6$\pm$0.3 & 15.7$\pm$2.3 & 687.8$\pm$2.1 & &  6.8$\pm$0.9 &
74$\pm$35 & & 325.1$\pm$11.5 \\
8.5$\pm$0.1 & 8.1$\pm$0.3 & 723.0$\pm$0.9 & &  5.6$\pm$0.7 & 83$\pm$31
& & 329.5$\pm$8.9 \\
8.7$\pm$0.1 & 7.2$\pm$0.7  & 769.2$\pm$1.2  &  & 5.1$\pm$0.9 &
139$\pm$62 & & 288.6$\pm$31 \\

\end{tabular}
\end{center}
{\footnotesize $^a$ The frequency of the lower kHz QPO
($\nu_{lower~kHz}$) and all the parameters on the first row were
determined without shifting the data.}  

\end{table*}

\section{Discussion}
\noindent
We found that in 4U~1636--53 the kHz QPO frequency separation
($\Delta\nu$) changes as a function of the frequency of the lower kHz
QPO from 323.3$\pm$4.3 Hz to 242.4$\pm$3.6 Hz (Fig.~\ref{delta} {\em
left}); the maximum peak separation is significantly larger than 290.5
Hz (half the burst oscillation frequency of $\sim$581 Hz; see
Section~\ref{intro}), while the minimum peak separation is
significantly smaller than that. The change in $\Delta\nu$ as a
function of upper kHz QPO frequency is abrupt; the data are
consistent with a sudden jump at $\nu=1050$~Hz (Fig.~\ref{delta} {\em
right}). We note that similar jumps in $\Delta\nu$ as a function of
the upper kHz QPO have also been seen in 4U~1728--34 and 4U~1608--52,
although in those cases $\Delta\nu$ always remained smaller or equal
to (half) the burst oscillation frequency
(\pcite{1999ApJ...517L..51M}; \pcite{mevawi1998};
\pcite{2001ApJ...554.1210L}; \pcite{2001ApJ...553L.157M}). This is the
first time that the kHz peak separation has been shown to be
significantly larger than the inferred neutron star spin frequency and
also the first time that $\Delta\nu$ has been seen to vary between
less and more than (half) the burst oscillation frequency.

In previous work burst oscillations detected in the dipping burster
4U~1916--05 at $\sim$272 Hz also seemed to occur at too low a
frequency for $\Delta\nu$ (\pcite{2001ApJ...549L..85G}). However, in
that source the kHz QPOs are weak and sometimes very broad, therefore
whole segments of several ksec of data had to be averaged to obtain a
detection (\pcite{bobaol2000}). This means that there is no certainty
that the kHz QPOs were detected simultaneously or that their apparent
separation was not affected by frequency shifts combined with changes
in their power ratio. Therefore, the measured $\Delta\nu$ may not
reflect the true peak separation. In the current work, using the
shift--and--add method we excluded the possibility that effects
changing $\Delta\nu$ on timescales longer than 256~s have influenced
the observed $\Delta\nu$. On timescales shorter than 256~s, changes in
the rms amplitudes could bias the peak separation for reasons outlined
above. However, since our measurements show that the lower kHz QPO rms
amplitude increased while the rms amplitude of the upper kHz QPO
decreased this effect will have made the observed $\Delta\nu$ smaller
than the true peak separation assuming the rms amplitude--frequency
relation is the same for timescales shorter than 256~s. Fast changes
in the upper kHz QPO frequency, associated with the 5--6 Hz QPO in
Sco~X--1 ($\sim$20 Hz; \pcite{2001ApJ...559L..29Y}) could also bias
$\Delta\nu$ measurements. However, those $\sim$5--6 Hz QPOs are only
found when atoll and Z sources are in a high--luminosity state
(e.g. \pcite{1986ApJ...306..230M};
\pcite{1999ApJ...512L..39W}). Changes in the kHz QPO frequencies
associated with milliHz QPOs present in 4U~1636--53 and 4U~1608--52 in
a small luminosity interval (\pcite{2001A&A...372..138R}) are so small
(less than a Hz; \pcite{2002ApJ...567L..67Y}) that it is unlikely that
they alter our findings significantly, although we note that such a
milliHz QPO was present during some of the observations.

Systematic variations in the profile of the kHz QPOs could in
principle cause the kHz QPO separation frequency to become large at
low kHz QPO frequencies and small at high frequencies since the
centroid frequencies have been measured assuming a Lorentzian peak
profile. Since our fits were good (reduced $\chi^2\sim$1) and no
systematic residuals were apparent, we estimate that shifts in the
centroid frequency due to asymmetries in the peak profiles could at
most be as large as the FWMH of the peak divided by its significance,
i.e. negligible in case of the lower kHz QPO but up to $\sim$10~Hz in
case of the upper kHz QPO. For power spectra with a lower kHz QPO in
the frequency range 650--750 Hz, the average data$-$model residuals
within one FWHM are less then 0.15 per cent of the power in the
peak. Furthermore, there is no evidence for a sudden change of the QPO
profile which could produce the observed abrupt decrease in
$\Delta\nu$. When using Gaussians to model the kHz QPO peaks the fits
were also good (reduced $\chi^2\sim$1) and we obtained results
consistent with those using Lorentzians. We conclude that it is
unlikely that our results can be explained by changing asymmetries in
the peak profile of the kHz QPOs.

The principal motivation for a beat frequency model is the relative
closeness of the inferred spin frequency to $\Delta\nu$. We found that
$\Delta\nu$ is consistent with being distributed symmetrically around
half the burst oscillation frequency, emphasizing the apparent
connection between the inferred spin and $\Delta\nu$. The present
formulation of the sonic--point beat--frequency model can explain the
observed decrease in $\Delta\nu$ as a function of kHz QPO frequency
(\pcite{2001ApJ...554.1210L}). However, all versions of the
sonic--point beat--frequency model described so far involving a
prograde spinning accretion disk (\pcite{milaps1998};
\pcite{2001ApJ...554.1210L}) predict that $\Delta\nu$ is always less
than or equal to the spin frequency of the neutron star for Keplerian
orbital frequencies larger than the neutron star spin frequencies;
since our results for 4U~1636--53 show that $\Delta\nu$ can be larger
than the neutron star spin frequency this prediction is now falsified.

\scite{2001AdSpR..28..511S} showed that it is still unclear whether
the burst oscillations at $\sim$581~Hz reflect the spin frequency of
the neutron star or its second harmonic as suggested by
\scite{mi1999}. Therefore, one might argue that the spin frequency is
really at 581~Hz (not 290.5~Hz) and hence that $\Delta\nu$ is less
than the spin frequency. This would, however, take away the principal
motivation of beat--frequency models, since the neutron star spin
frequency and the kHz QPO peak separation are in that case completely
different.

That in the sonic--point beat--frequency model the neutron star spin
and $\Delta\nu$ are not exactly equal is explained by the way in which
the beat--frequency interaction produces the observed frequencies. The
key element is the radial motion of the plasma clumps near the sonic
radius (\pcite{2001ApJ...554.1210L}). This makes the observed lower
kHz QPO frequency somewhat larger than the beat--frequency since it
squeezes the spatial separation between peaks of enhanced mass
flow. It also makes the upper kHz QPO frequency smaller than the
orbital frequency since it makes the clump's footpoint move backward
in a corotating frame. It might be possible to find other effects in
the plasma flow patterns besides those discussed by
\scite{2001ApJ...554.1210L} (all of which make $\Delta\nu$ smaller)
which may lead to larger values of $\Delta\nu$ at lower kHz QPO
frequencies. Fully relativistic simulations of the gas flow and
radiation transport reported by \scite{2001ApJ...554.1210L} have not,
so far, shown evidence for this. If, however, the sense of rotation of
(part of) the accretion disk would change from prograde to retrograde
when $\Delta\nu$ changes from less to more than the spin frequency,
then our results could still be in accordance with current
formulations of the sonic--point beat--frequency model. This also
requires that in our observations for $\Delta\nu$ larger than 290.5~Hz
the lower kHz QPO reflects the (near) Keplerian orbital frequency,
whereas for those points in Figure~\ref{delta} where $\Delta\nu$ is
smaller than half the burst oscillation frequency, the Keplerian
orbital motion is represented by the upper kHz QPO.

We detected a low--frequency QPO between $\sim$25--40 Hz in
4U~1636--53 simultaneously with the kHz QPO pair. The relation between
this low--frequency QPO and the lower kHz QPO is similar to the one
found by \scite{fova1998} for 4U~1728--34. It has been suggested that
such low--frequency QPOs in atoll sources are related to the Z source
Horizontal Branch Oscillations (HBO; e.g. \pcite{1998ApJ...499L..41H};
\pcite{psbeva1999}). According to the magnetospheric beat--frequency
model these HBO originate near the magnetospheric radius
(\pcite{alsh1985}; \pcite{lashal1985}). The frequency of the
low--frequency QPO in 4U~1636--53 increased as the frequency of the
lower kHz QPO increased. Therefore, if the disk near the sonic radius
is indeed retrograde within our observations the retrograde annulus
must extend at least up to the magnetospheric radius if the
sonic--point and magnetospheric beat--frequency model both
apply. Partially retrograde spinning disks have been
proposed as an explanation for observed torque reversals
(\pcite{1997ApJS..113..367B}) in slowly rotating neutron stars with a
dipole magnetic field strength of 10$^{11-12}$ Gauss
(\pcite{1998ApJ...499L..27V}). Modelling by
\scite{1998ApJ...499L..27V} shows that the accretion disk can get
inclinations of more than 90 degrees. It is not clear if their work is
also applicable to the case of low--magnetic field LMXBs. Clearly,
allowing disks with alternating rotation senses complicates the
sonic--point beat frequency interpretation of the kHz QPOs
considerably.

\section*{Acknowledgments} 
\noindent 
PGJ is supported by EC Marie Curie Fellowship
HPMF--CT--2001--01308. MK is supported in part by a Netherlands
Organization for Scientific Research (NWO) grant. We would like to
thank Cole Miller for useful discussions and the referee for comments
which helped improve the manuscript.

\end{document}